\input harvmac
\noblackbox

\input epsf


\def\journal#1&#2(#3){\unskip, \sl #1\ \bf #2 \rm(19#3) }
\def\andjournal#1&#2(#3){\sl #1~\bf #2 \rm (19#3) }

\def\frac#1#2{{#1\over#2}}

\def\d{\partial}

\def\inbar{\,\vrule height1.5ex width.4pt depth0pt}
\def\IC{\relax\hbox{$\inbar\kern-.3em{\rm C}$}}
\def\IR{\relax{\rm I\kern-.18em R}}
\def\IP{\relax{\rm I\kern-.18em P}}
\def\IZ{\relax{\rm I\kern-.18em Z}}

%
%

%
\catcode`\@=11
\def\slash#1{\mathord{\mathpalette\c@ncel{#1}}}
\overfullrule=0pt

\def\DD{{\cal D}}

\def\FF{{\cal F}}

\def\II{{\cal I}}

\def\NN{{\cal N}}
\def\OO{{\cal O}}

\def\RR{{\cal R}}
\def\SS{{\cal S}}

\def\VV{{\cal V}}

\def\YY{{\cal Y}}
\def\ZZ{{\cal Z}}

\def\underrel#1\over#2{\mathrel{\mathop{\kern\z@#1}\limits_{#2}}}

\catcode`\@=12


%



\def\unlockat{\catcode`\@=11}
\def\lockat{\catcode`\@=12}

\unlockat


\def\newsec#1{\global\advance\secno by1\message{(\the\secno. #1)}
\global\subsecno=0\global\subsubsecno=0\eqnres@t\noindent
{\bf\the\secno. #1}
\writetoca{{\secsym} {#1}}\par\nobreak\medskip\nobreak}
\global\newcount\subsecno \global\subsecno=0
\def\subsec#1{\global\advance\subsecno
by1\message{(\secsym\the\subsecno. #1)}
\ifnum\lastpenalty>9000\else\bigbreak\fi\global\subsubsecno=0
\noindent{\it\secsym\the\subsecno. #1}
\writetoca{\string\quad {\secsym\the\subsecno.} {#1}}
\par\nobreak\medskip\nobreak}
\global\newcount\subsubsecno \global\subsubsecno=0
\def\subsubsec#1{\global\advance\subsubsecno by1
\message{(\secsym\the\subsecno.\the\subsubsecno. #1)}
\ifnum\lastpenalty>9000\else\bigbreak\fi
\noindent\quad{\secsym\the\subsecno.\the\subsubsecno.}{#1}
\writetoca{\string\qquad{\secsym\the\subsecno.\the\subsubsecno.}{#1}}
\par\nobreak\medskip\nobreak}

\def\subsubseclab#1{\DefWarn#1\xdef
#1{\noexpand\hyperref{}{subsubsection}%
{\secsym\the\subsecno.\the\subsubsecno}%
{\secsym\the\subsecno.\the\subsubsecno}}%
\writedef{#1\leftbracket#1}\wrlabeL{#1=#1}}
\lockat


\newcount\figno
\figno=0
\def\fig#1#2#3{
\par\begingroup\parindent=0pt\leftskip=1cm\rightskip=1cm\parindent=0pt
\baselineskip=11pt
\global\advance\figno by 1
\midinsert
\epsfxsize=#3
\centerline{\epsfbox{#2}}
\vskip 12pt
{\bf Fig.\ \the\figno: } #1\par
\endinsert\endgroup\par
}
\def\figlabel#1{\xdef#1{\the\figno}}
\def\encadremath#1{\vbox{\hrule\hbox{\vrule\kern8pt\vbox{\kern8pt
\hbox{$\displaystyle #1$}\kern8pt}
\kern8pt\vrule}\hrule}}
%
%


\font\cmss=cmss10
\font\cmsss=cmss10 at 7pt
\def\rlx{\relax\leavevmode}
\def\inbar{\vrule height1.5ex width.4pt depth0pt}
\def\IC{\relax\,\hbox{$\inbar\kern-.3em{\rm C}$}}
\def\IN{\relax{\rm I\kern-.18em N}}
\def\IP{\relax{\rm I\kern-.18em P}}
\def\ZZ{\rlx\leavevmode\ifmmode\mathchoice{\hbox{\cmss Z\kern-.4em Z}}
 {\hbox{\cmss Z\kern-.4em Z}}{\lower.9pt\hbox{\cmsss Z\kern-.36em Z}}
 {\lower1.2pt\hbox{\cmsss Z\kern-.36em Z}}\else{\cmss Z\kern-.4em
 Z}\fi}
\def\IZ{\relax\ifmmode\mathchoice
{\hbox{\cmss Z\kern-.4em Z}}{\hbox{\cmss Z\kern-.4em Z}}
{\lower.9pt\hbox{\cmsss Z\kern-.4em Z}}
{\lower1.2pt\hbox{\cmsss Z\kern-.4em Z}}\else{\cmss Z\kern-.4em
Z}\fi}
\def\IZ{\relax\ifmmode\mathchoice
{\hbox{\cmss Z\kern-.4em Z}}{\hbox{\cmss Z\kern-.4em Z}}
{\lower.9pt\hbox{\cmsss Z\kern-.4em Z}}
{\lower1.2pt\hbox{\cmsss Z\kern-.4em Z}}\else{\cmss Z\kern-.4em
Z}\fi}

\def\narrowplus{\kern -.04truein + \kern -.03truein}
\def\narrowminus{- \kern -.04truein}
\def\narrowminussub{\kern -.02truein - \kern -.01truein}

\def\IZ{\relax\ifmmode\mathchoice
{\hbox{\cmss Z\kern-.4em Z}}{\hbox{\cmss Z\kern-.4em Z}}
{\lower.9pt\hbox{\cmsss Z\kern-.4em Z}}
{\lower1.2pt\hbox{\cmsss Z\kern-.4em Z}}\else{\cmss Z\kern-.4em
Z}\fi}
\def\IB{\relax{\rm I\kern-.18em B}}
\def\IC{{\relax\hbox{$\inbar\kern-.3em{\rm C}$}}}
\def\ID{\relax{\rm I\kern-.18em D}}
\def\IE{\relax{\rm I\kern-.18em E}}
\def\IF{\relax{\rm I\kern-.18em F}}
\def\IG{\relax\hbox{$\inbar\kern-.3em{\rm G}$}}
\def\IGa{\relax\hbox{${\rm I}\kern-.18em\Gamma$}}
\def\IH{\relax{\rm I\kern-.18em H}}
\def\II{\relax{\rm I\kern-.18em I}}
\def\IK{\relax{\rm I\kern-.18em K}}
\def\IP{\relax{\rm I\kern-.18em P}}

\font\cmss=cmss10 \font\cmsss=cmss10 at 7pt
\def\IR{\relax{\rm I\kern-.18em R}}

\def\dab#1{ { \partial \over \partial #1} }


%

%
%
\def\eqnn#1{\xdef #1{(\secsym\the\meqno)}\writedef{#1\leftbracket#1}%
\global\advance\meqno by1\wrlabeL#1}
\def\eqna#1{\xdef #1##1{\hbox{$(\secsym\the\meqno##1)$}}
\writedef{#1\numbersign1\leftbracket#1{\numbersign1}}%
\global\advance\meqno by1\wrlabeL{#1$\{\}$}}
\def\eqn#1#2{\xdef #1{(\secsym\the\meqno)}\writedef{#1\leftbracket#1}%
\global\advance\meqno by1$$#2\eqno#1\eqlabeL#1$$}


\def\boxit#1{\vbox{\hrule\hbox{\vrule\kern8pt
\vbox{\hbox{\kern8pt}\hbox{\vbox{#1}}\hbox{\kern8pt}}
\kern8pt\vrule}\hrule}}
\def\mathboxit#1{\vbox{\hrule\hbox{\vrule\kern5pt\vbox{\kern5pt
\hbox{$\displaystyle #1$}\kern5pt}\kern5pt\vrule}\hrule}}


\lref\KutasovPV{
D.~Kutasov,
``Some properties of (non)critical strings,''
arXiv:hep-th/9110041.
}

\lref\MukhiZB{
S.~Mukhi and C.~Vafa,
``Two-dimensional black hole as a topological coset model of c = 1 string
theory,''
Nucl.\ Phys.\ B {\bf 407}, 667 (1993)
[arXiv:hep-th/9301083].
}

\lref\DijkgraafHK{
R.~Dijkgraaf, G.~W.~Moore and R.~Plesser,
``The Partition function of 2-D string theory,''
Nucl.\ Phys.\ B {\bf 394}, 356 (1993)
[arXiv:hep-th/9208031].
}

\lref\AharonyVK{
O.~Aharony, B.~Fiol, D.~Kutasov and D.~A.~Sahakyan,
``Little string theory and heterotic/type II duality,''
Nucl.\ Phys.\ B {\bf 679}, 3 (2004)
[arXiv:hep-th/0310197].
}

\lref\TakayanagiYB{
  T.~Takayanagi,
  ``$c < 1$ string from two dimensional black holes,''
  arXiv:hep-th/0503237.
}

\lref\RibaultWP{
  S.~Ribault and J.~Teschner,
  ``H(3)+ correlators from Liouville theory,''
  arXiv:hep-th/0502048.
}

\lref\RibaultMS{
  S.~Ribault,
  ``Knizhnik-Zamolodchikov equations and spectral flow in AdS3 string theory,''
  arXiv:hep-th/0507114.
}

\lref\GiribetIX{
  G.~Giribet and Y.~Nakayama,
  ``The Stoyanovsky-Ribault-Teschner map and string scattering amplitudes,''
  arXiv:hep-th/0505203.
}

\lref\AganagicQJ{
  M.~Aganagic, R.~Dijkgraaf, A.~Klemm, M.~Marino and C.~Vafa,
  ``Topological strings and integrable hierarchies,''
  arXiv:hep-th/0312085.
}

\lref\GhoshalWM{
  D.~Ghoshal and C.~Vafa,
  ``C = 1 string as the topological theory of the conifold,''
  Nucl.\ Phys.\ B {\bf 453}, 121 (1995)
  [arXiv:hep-th/9506122].
}

\lref\OoguriWJ{
  H.~Ooguri and C.~Vafa,
  ``Two-Dimensional Black Hole and Singularities of CY Manifolds,''
  Nucl.\ Phys.\ B {\bf 463}, 55 (1996)
  [arXiv:hep-th/9511164].
}

\lref\nania{
S.~Nakamura and V.~Niarchos,
work in progress.
}

\lref\StoyanovskyPG{
  A.~V.~Stoyanovsky,
  ``A relation between the Knizhnik--Zamolodchikov and
  Belavin--Polyakov--Zamolodchikov systems of partial differential equations,''
  arXiv:math-ph/0012013.
}

\lref\AharonyXN{
  O.~Aharony, A.~Giveon and D.~Kutasov,
  ``LSZ in LST,''
  Nucl.\ Phys.\ B {\bf 691}, 3 (2004)
  [arXiv:hep-th/0404016].
}

\lref\TakayanagiYR{
T.~Takayanagi,
``Matrix model and time-like linear dilaton matter,''
JHEP {\bf 0412}, 071 (2004)
[arXiv:hep-th/0411019].
}

\lref\MaldacenaHW{
  J.~M.~Maldacena and H.~Ooguri,
  ``Strings in AdS(3) and SL(2,R) WZW model. I,''
  J.\ Math.\ Phys.\  {\bf 42}, 2929 (2001)
  [arXiv:hep-th/0001053].
}

\lref\DijkgraafDJ{
  R.~Dijkgraaf, H.~L.~Verlinde and E.~P.~Verlinde,
  ``Topological Strings In $D < 1$,''
  Nucl.\ Phys.\ B {\bf 352}, 59 (1991).
}

\lref\WarnerZH{
  N.~P.~Warner,
  ``N=2 supersymmetric integrable models and topological field theories,''
  arXiv:hep-th/9301088.
}

\lref\WittenZZ{
  E.~Witten,
  ``Mirror manifolds and topological field theory,''
  arXiv:hep-th/9112056.
}

\lref\KlebanovQA{
  I.~R.~Klebanov,
  ``String theory in two-dimensions,''
  arXiv:hep-th/9108019.
}

\lref\GinspargIS{
  P.~H.~Ginsparg and G.~W.~Moore,
  ``Lectures on 2-D gravity and 2-D string theory,''
  arXiv:hep-th/9304011.
}

\lref\PolchinskiMB{
  J.~Polchinski,
  ``What is string theory?,''
  arXiv:hep-th/9411028.
}

\lref\WittenXJ{
  E.~Witten,
  ``Topological Sigma Models,''
  Commun.\ Math.\ Phys.\  {\bf 118}, 411 (1988).
}

\lref\GrossAY{
  D.~J.~Gross and N.~Miljkovic,
  ``A Nonperturbative Solution Of D = 1 String Theory,''
  Phys.\ Lett.\ B {\bf 238}, 217 (1990).
}

\lref\BrezinSS{
  E.~Brezin, V.~A.~Kazakov and A.~B.~Zamolodchikov,
  ``Scaling Violation In A Field Theory Of Closed Strings In One Physical
  Dimension,''
  Nucl.\ Phys.\ B {\bf 338}, 673 (1990).
}

\lref\GinspargAS{
  P.~H.~Ginsparg and J.~Zinn-Justin,
  ``2-D Gravity + 1-D Matter,''
  Phys.\ Lett.\ B {\bf 240}, 333 (1990).
}

\lref\McGreevyKB{
  J.~McGreevy and H.~L.~Verlinde,
  ``Strings from tachyons: The c = 1 matrix reloaded,''
  JHEP {\bf 0312}, 054 (2003)
  [arXiv:hep-th/0304224].
}

\lref\McGreevyEP{
  J.~McGreevy, J.~Teschner and H.~L.~Verlinde,
  ``Classical and quantum D-branes in 2D string theory,''
  JHEP {\bf 0401}, 039 (2004)
  [arXiv:hep-th/0305194].
}

\lref\KlebanovKM{
  I.~R.~Klebanov, J.~Maldacena and N.~Seiberg,
  ``D-brane decay in two-dimensional string theory,''
  JHEP {\bf 0307}, 045 (2003)
  [arXiv:hep-th/0305159].
}

\lref\SenIV{
  A.~Sen,
  ``Open-closed duality: Lessons from matrix model,''
  Mod.\ Phys.\ Lett.\ A {\bf 19}, 841 (2004)
  [arXiv:hep-th/0308068].
}

\lref\GaiottoYB{
  D.~Gaiotto and L.~Rastelli,
  ``A paradigm of open/closed duality: Liouville D-branes and the  Kontsevich
  model,''
  arXiv:hep-th/0312196.
}

\lref\OhtaEH{
  N.~Ohta and H.~Suzuki,
  ``Bosonization of a topological coset model and noncritical string theory,''
  Mod.\ Phys.\ Lett.\ A {\bf 9}, 541 (1994)
  [arXiv:hep-th/9310180].
}

\lref\GhoshalQT{
  D.~Ghoshal and S.~Mukhi,
  ``Topological Landau-Ginzburg model of two-dimensional string theory,''
  Nucl.\ Phys.\ B {\bf 425}, 173 (1994)
  [arXiv:hep-th/9312189].
}

\lref\VafaWI{
  C.~Vafa,
  ``Superstrings and topological strings at large N,''
  J.\ Math.\ Phys.\  {\bf 42}, 2798 (2001)
  [arXiv:hep-th/0008142].
}

\lref\WittenYJ{
  E.~Witten and B.~Zwiebach,
  ``Algebraic structures and differential geometry in $2-D$ string theory,''
  Nucl.\ Phys.\ B {\bf 377}, 55 (1992)
  [arXiv:hep-th/9201056].
}

\lref\DiFrancescoUD{
  P.~Di Francesco and D.~Kutasov,
  ``World sheet and space-time physics in two-dimensional (Super)string
  theory,''
  Nucl.\ Phys.\ B {\bf 375}, 119 (1992)
  [arXiv:hep-th/9109005].
}

\lref\LosevTT{
  A.~Losev,
  ``Descendants constructed from matter field in topological Landau-Ginzburg
  theories coupled to topological gravity,''
  Theor.\ Math.\ Phys.\  {\bf 95}, 595 (1993)
  [Teor.\ Mat.\ Fiz.\  {\bf 95}, 307 (1993)]
  [arXiv:hep-th/9211090].
}

\lref\EguchiXX{
  T.~Eguchi, H.~Kanno, Y.~Yamada and S.~K.~Yang,
  ``Topological strings, flat coordinates and gravitational descendants,''
  Phys.\ Lett.\ B {\bf 305}, 235 (1993)
  [arXiv:hep-th/9302048].
}

\lref\LercheUY{
  W.~Lerche, C.~Vafa and N.~P.~Warner,
  ``Chiral Rings In N=2 Superconformal Theories,''
  Nucl.\ Phys.\ B {\bf 324}, 427 (1989).
}

\lref\GiveonZM{
  A.~Giveon, D.~Kutasov and O.~Pelc,
  ``Holography for non-critical superstrings,''
  JHEP {\bf 9910}, 035 (1999)
  [arXiv:hep-th/9907178].
}

\lref\OoguriFP{
  H.~Ooguri and C.~Vafa,
  ``Geometry of N=2 strings,''
  Nucl.\ Phys.\ B {\bf 361}, 469 (1991).
}

\lref\OoguriIE{
  H.~Ooguri and C.~Vafa,
  ``N=2 heterotic strings,''
  Nucl.\ Phys.\ B {\bf 367}, 83 (1991).
}

\lref\AharonyVK{
  O.~Aharony, B.~Fiol, D.~Kutasov and D.~A.~Sahakyan,
  ``Little string theory and heterotic/type II duality,''
  Nucl.\ Phys.\ B {\bf 679}, 3 (2004)
  [arXiv:hep-th/0310197].
}

\lref\KonechnyDF{
  A.~Konechny, A.~Parnachev and D.~A.~Sahakyan,
  ``The ground ring of N = 2 minimal string theory,''
  arXiv:hep-th/0507002.
}

\lref\TakayanagiYJ{
  T.~Takayanagi,
  ``Notes on S-matrix of non-critical N = 2 string,''
  arXiv:hep-th/0507065.
}

\lref\GatoRiveraKI{
  B.~Gato-Rivera and A.~M.~Semikhatov,
  ``Minimal models from W constrained hierarchies via the Kontsevich-Miwa
  transform,''
  Phys.\ Lett.\ B {\bf 288}, 38 (1992)
  [arXiv:hep-th/9204085].
}

\lref\BershadskyQG{
  M.~Bershadsky, W.~Lerche, D.~Nemeschansky and N.~P.~Warner,
  ``A BRST operator for noncritical W strings,''
  Phys.\ Lett.\ B {\bf 292}, 35 (1992)
  [arXiv:hep-th/9207067].
}

\lref\KutasovUF{
  D.~Kutasov,
  ``Introduction to little string theory,''
{\it Prepared for ICTP Spring School on Superstrings and Related Matters, Trieste, Italy, 2-10 Apr 2001}
}

\lref\AharonyKS{
  O.~Aharony,
  ``A brief review of 'little string theories',''
  Class.\ Quant.\ Grav.\  {\bf 17}, 929 (2000)
  [arXiv:hep-th/9911147].
}

\lref\DijkgraafFC{
  R.~Dijkgraaf and C.~Vafa,
  ``Matrix models, topological strings, and supersymmetric gauge theories,''
  Nucl.\ Phys.\ B {\bf 644}, 3 (2002)
  [arXiv:hep-th/0206255].
}

\lref\WittenFB{
  E.~Witten,
  ``Chern-Simons gauge theory as a string theory,''
  Prog.\ Math.\  {\bf 133}, 637 (1995)
  [arXiv:hep-th/9207094].
}

\lref\HoriAX{
  K.~Hori and A.~Kapustin,
  ``Duality of the fermionic 2d black hole and N = 2 Liouville theory as
  mirror symmetry,''
  JHEP {\bf 0108}, 045 (2001)
  [arXiv:hep-th/0104202].
}



\Title{
}
{\vbox{\centerline{Notes On The S-Matrix Of Bosonic And}
\vskip 10pt \centerline{Topological Non-Critical Strings}
}}
\bigskip
\centerline{Shin Nakamura\footnote{$^{\dagger}$}{nakamura@nbi.dk}
and Vasilis Niarchos\footnote{$^{\ddagger}$}{niarchos@nbi.dk}}
\bigskip
\centerline{{\it The Niels Bohr Institute}}
\centerline{\it Blegdamsvej 17, 2100 Copenhagen \O, Denmark}
\bigskip\bigskip\bigskip
\noindent

We show that the equivalence between the $c=1$ non-critical
bosonic string and the $\NN=2$ topologically twisted coset
$SL(2)/U(1)$ at level one can be checked very naturally on the level
of tree-level scattering amplitudes with the use of the
Stoyanovsky-Ribault-Teschner map, which recasts $H_3^+$
correlation functions in terms of Liouville field theory
amplitudes. This observation can be applied
equally well to the topologically twisted
$SL(2)_n/U(1)$ coset with $n>1$, which has been argued recently
to be equivalent with a $c<1$ non-critical bosonic string
whose matter part is defined by a time-like linear dilaton
CFT.

\vfill
\Date{July 2005}


\listtoc
\writetoc

\newsec{Introduction}

String theories in two dimensions \refs{\KlebanovQA\GinspargIS-\PolchinskiMB}
and $\NN=2$ topological strings \refs{\WittenXJ,\WittenZZ}
have been studied over the years in much detail.
In both cases the dynamics are governed by some integrable
structure. Indeed, two dimensional non-critical strings have been
re-formulated and solved in many cases in terms of
a dual matrix model
\refs{\GrossAY\BrezinSS-\GinspargAS,\KlebanovQA\GinspargIS-\PolchinskiMB}.
The emergence of the matrix model
has been understood recently in terms of a holographic open/closed
string duality \refs{\McGreevyKB\McGreevyEP\KlebanovKM-\SenIV}.
Similarly, $\NN=2$ topological strings
on generic non-compact Calabi-Yau spaces have been argued to be
dual to corresponding finite $N$ or large $N$ matrix models
through a rather different class of holographic open/closed string
dualities \refs{\WittenFB\DijkgraafFC\AganagicQJ-\GaiottoYB}.
All these features are interesting and provide a useful framework,
where many aspects of string theory and quantum gravity
can be studied efficiently.

Because of the presence of a topological/integrable
structure in both two-dimensional and topological string theories,
it is perhaps not very surprising that
we can find examples, where a direct connection exists between
a two dimensional non-critical string theory and a topological
string theory on some appropriate target space.
An example of this sort of equivalence
involves the $c=1$ non-critical bosonic string theory at the self-dual radius
and the topologically twisted $\NN=2$ coset CFT $SL(2,\IR)_1/U(1)$
at level one \MukhiZB. Strong evidence in favor
of this equivalence has been presented
on the level of BRST cohomologies
and correlation functions in \refs{\MukhiZB,\OhtaEH}.
It has been further
argued that there is another equivalence between
the topologically twisted coset
$SL(2,\IR)_1/U(1)$ and the topological
Landau-Ginzburg (LG) model with superpotential $W(X)=X^{-1}$ \GhoshalQT.
This suggests an additional connection between the topologically
twisted coset CFT and the topological string theory on the conifold
\refs{\GhoshalWM,\OoguriWJ}. The existence of the same set of deformations
on both sides of this equivalence have led to the intriguing
claim that the $\NN=2$ topological string on generic Calabi-Yau
threefold singularities is described by appropriate deformations of the
$c=1$ non-critical bosonic string theory \VafaWI.

One may wonder whether
it is possible to find bosonic string equivalents to the topologically
twisted coset $SL(2)_n/U(1)$ at general values of the level $n$.
It turns out that the answer to this question is affirmative.
The corresponding bosonic string has been determined recently in \TakayanagiYB.
For integer levels $n$ the precise statement
is that the A-model topological string on the $\IZ_n$ orbifold
$\bigg( \frac{SL(2,\IR)_n}{U(1)}\bigg)/\IZ_n$
is equivalent to the non-minimal $c=1-6(n-1)^2/n$
non-critical bosonic string. The definition
of the non-minimal $c=1-6(n-1)^2/n$ bosonic string
involves the standard Liouville sector with linear
dilaton slope $Q=\frac{1}{\sqrt{n}}(n+1)$ plus
a time-like boson $X_0$ that is compactified
at radius $R_{X_0}=\sqrt{2n}$ and has linear
dilaton slope $q=\frac{1}{\sqrt{n}}(n-1)$.
As expected, for $n=1$ the non-minimal string reduces
to the usual $c=1$ bosonic string and we recover
the equivalence of the previous paragraph.
Moreover, as in the case of the $c=1$ string, these theories are
expected to be equivalent to topological LG models
with a more general superpotential $W(X)=-\mu X^{-n}$.
In \TakayanagiYB\ Takayanagi supported the claim
for this general $n$ equivalence with an
analysis of the respective BRST cohomologies,
a computation of tree-level two- and three-point functions in
the topologically twisted $SL(2,\IR)_n/U(1)$
coset and a computation of higher $N$-point functions
using topological LG techniques.

In this note, we provide further evidence for this equivalence
with a computation of a large class of tree-level
$N$-point correlation functions directly
in the A-model topological string on
$\bigg( \frac{SL(2,\IR)_n}{U(1)}\bigg)/\IZ_n$.
At first sight, this computation appears to be a formidable
task that would require explicit knowledge of
arbitrary $N$-point functions of the $SL(2,\IR)/U(1)$
theory. Fortunately, however, it turns out that
we can establish the agreement of the sphere correlation
functions on both sides of the equivalence in
a rather simple and straightforward way.
This new non-trivial check relies heavily
on work by Ribault and Teschner that appeared
recently in two beautiful papers \refs{\RibaultWP,\RibaultMS}.
In the first of these papers,
an intriguing new dictionary was established between
tree-level $N$-point functions in the $SL(2,\IC)/SU(2)$ WZW model
at level $n$ and $(2N-2)$-point functions in Liouville field theory.
This correspondence is based on a map between solutions of the
Belavin-Polyakov-Zamolodchikov (BPZ)
and Knizhnik-Zamolodchikov (KZ) systems of partial differential
equations that was discovered originally by Stoyanovsky in \StoyanovskyPG.
In a second paper \RibaultMS\ this dictionary
was extended to include also spectral-flow violating amplitudes
in $SL(2,\IR)$ (see also related work in \GiribetIX\foot{In \GiribetIX\
one can also find a nice discussion on the implications of the
Stoyanovsky-Ribault-Teschner map in Little String Theory.}).
The $SL(2,\IC)/SU(2)$ WZW model is the Euclidean version
of the $SL(2,\IR)$ WZW model and we will go from one to the other
by analytic continuation. Partial justification for the
validity of this continuation will be given by the consistency of
our results.

To summarize, our strategy is the following.
We begin with a certain class of sphere $N$-point correlation
functions in the topologically twisted $SL(2,\IR)/U(1)$ theory.
Then, we show that these correlation functions
can be reduced easily to a set of $SL(2,\IR)$
amplitudes with maximal spectral-flow number violation
and we make use of the Stoyanovsky-Ribault-Teschner
(SRT) map to recast these amplitudes
in terms of Liouville field theory $N$-point functions.
Through this map one can see
the non-minimal bosonic string amplitudes emerging
naturally with the right parameters and with precisely
those insertions expected from the bosonic/topological
correspondence of \TakayanagiYB.
One of the most appealing features of this approach is that it
gives a very natural and straightforward way
to go from the topological string to the non-critical bosonic
string. We believe that similar techniques will be useful
in the study of topological strings in many other cases
where the $SL(2)/U(1)$ coset is involved. This includes
topological strings in generalized conifold singularities in
CY threefolds and the $\NN=2$ string (equivalently the $\NN=4$ topological
string) in the vicinity of K3 singularities. We make
several comments on such generalizations in the final
section.

The organization of this paper is as follows. In section
2 we review the basic features of the equivalence proposed in
\TakayanagiYB\ and set our notation straight. In section 3
we present the SRT map, which will be
used heavily in section 4 to compute
topological string $N$-point functions.
We conclude in section 5 with a brief
discussion of some interesting open problems.

\newsec{Topologically twisted 2D black holes and bosonic
non-critical strings}

In this section we fix the notation and state in precise terms the
conjectured equivalence \TakayanagiYB\ between the $c\leq 1$ non-minimal
strings and the A-model topological string on
$\bigg( \frac{SL(2,\IR)_n}{U(1)}\bigg)/\IZ_n$.
We follow the notation of \TakayanagiYB\ and set $\alpha'=2$
throughout the rest of this note.

\subsec{The topologically twisted coset}

The $\NN=2$ supersymmetric $SL(2,\IR)_n/U(1)$ model at
level $n>0$ is a two-dimensional CFT with central charge
$c=3+\frac{6}{n}$. It can be obtained by gauging
an appropriate $U(1)$ in the product of
a free Dirac fermion $(\psi,\psi^{\dagger})$
and the $bosonic$ $SL(2,\IR)_{n+2}$
WZW model at level $n+2$. We summarize briefly
a few of the relevant details to fix the notation.

The bosonic $SL(2,\IR)_{n+2}$ WZW model involves an
$SL(2,\IR)_L\times SL(2,\IR)_R$ current algebra, whose left-moving
part is captured by the current OPE's
\eqn\aaa{
J^3(z)J^3(0)\sim -\frac{n+2}{2z^2}, ~ ~
J^3(z)J^{\pm}(0)\sim \pm \frac{J^{\pm}(0)}{z}, ~ ~
J^+(z)J^-(0)\sim \frac{n+2}{z^2}-\frac{2J^3(0)}{z}.
}
A similar set of OPE's determines the right-moving current algebra
$SL(2,\IR)_R$.
Now consider the product CFT of this $SL(2,\IR)_{n+2}$ WZW
model with a free CFT consisting of a complex fermion $(\psi,\psi^{\dagger})$.
As usual, the free OPE of the complex fermion $\psi$ is
$\psi(z)\psi^{\dagger}(0)\sim \frac{1}{z}$ and for later
convenience we bosonize the fermion with a boson $H$, so that
\eqn\aab{
\psi=e^{iH}~, ~ ~ \psi\psi^{\dagger}=i\d H
~.}
The $U(1)$ current, whose gauging defines the $\NN=2$ supersymmetric
coset, takes the form
\eqn\aac{
J_g=J^3-\psi^{\dagger}\psi-i\sqrt{\frac{n}{2}}\d X
~,}
where $X$ is a free $U(1)$ boson with the standard OPE
$X(z)X(0)\sim -\log z$. To perform the $U(1)$ gauging on
the BRST level, we can add a system of $c=-2$ ghosts $(\xi,\eta)$ and define
the BRST charge
\eqn\aad{
Q_{U(1)}=\int dz ~ \xi(z)J_g(z)
~.}

The result of this procedure is a theory with
$\NN=(2,2)$ worldsheet supersymmetry.
Geometrically, the target space of this theory looks like a cigar
(or the Euclidean version of a 2D black hole)
with asymptotic radius $R=\sqrt{2n}$.
The left-moving part of the $\NN=2$ superconformal current algebra
takes the form
\eqn\aae{\eqalign{
J_R&=\frac{n+2}{n}\psi^{\dagger}\psi-\frac{2}{n} J^3
\simeq -i\d H-i\sqrt{\frac{2}{n}}\d X~,
\cr
G^{+}&=\sqrt{\frac{2}{n}}\psi^{\dagger}J^+ ~, ~ ~
G^{-}=\sqrt{\frac{2}{n}}\psi J^-
~.}}
To obtain the second expression for the $U(1)_R$
current we used the $U(1)$ gauging condition.
Moreover, for later purposes, it will be
convenient to define an equivalent $U(1)_R$
current
\eqn\aaf{
J'_R=J_R-2J_g=-3i\d H -2 J^3 +i\sqrt{\frac{2}{n}}(n-1)\d X
~.}
Of course, similar statements apply also to the
right-movers, but with a few minor sign differences, which
we summarize here\foot{We use the following notation.
Bars $\bar{}$ will denote the right-movers and
daggers $^{\dagger}$ will denote complex conjugation
on the target fields.}
\eqn\aag{\eqalign{
\bar J_g&=-\bar J^3-\bar{\psi}{\bar \psi}^{\dagger}-
i\sqrt{\frac{n}{2}}\bar \d X~,
\cr
\bar J_R&=\frac{n+2}{n}\bar{\psi}{\bar \psi}^{\dagger}+
\frac{2}{n} \bar J^3\sim
\bar{\psi}{\bar \psi}^{\dagger}-i\sqrt{\frac{2}{n}}\bar \d X
~,
\cr
{\bar J}'_R&=\bar J_R-2\bar J_g=3\bar{\psi}{\bar \psi}^{\dagger}
+2 \bar J^3 +i\sqrt{\frac{2}{n}}(n-1)\bar \d X
~.}}

Certain primary fields of the supersymmetric
coset will soon play an important role.
They can be constructed from the bosonic $SL(2,\IR)_{n+2}$
primary fields, which we now quickly review.
The bosonic $SL(2,\IR)_{n+2}$
model possesses a set of affine primary fields
$\Phi_{j,m,\bar m}$ that
can be expressed as
\eqn\aal{
\Phi_{j,m,\bar m}=V_{j,m,\bar m}e^{\sqrt{\frac{2}{n+2}}(mX_3+\bar m \bar X_3)}
~.}
In this relation, $X_3$ is a canonically normalized
boson that has been defined so that
\eqn\aala{
J^3(z)=-\sqrt{\frac{n+2}{2}}\d X_3
~.}
Also, $V_{j,m,\bar m}$ is a primary field of the
bosonic coset $SL(2,\IR)_{n+2}/U(1)$
that carries by definition no $J^3_0$ charge.
The conformal dimensions of these fields are
\eqn\aam{
\Delta(\Phi_{j,m})=-\frac{j(j+1)}{n}~, ~ ~
\Delta(V_{j,m})=-\frac{j(j+1)}{n}+\frac{m^2}{n+2}
~.}
In these expressions the quantum
numbers $j$ and $m$ can take different values
depending on the type of $SL(2)$ representation
we want to consider. For the lowest weight discrete
representations $\DD^+_j$ $j$ is real
and $m=-j,-j+1,-j+2,...$. For the highest weight discrete
representations $\DD^-_j$ $j$ is again real, but
$m=j,j-1,j-2,...$. Finally, for the continuous representations
$j=-\frac{1}{2}+is$, $s\in \IR$, but
these representations will not appear in the
topologically twisted theory.

In addition, the spectrum of the bosonic $SL(2,\IR)$ WZW model includes
representations that arise from the spectral flow operation \MaldacenaHW,
some details of which are summarized in appendix A.
The spectral flow of the primary fields $\Phi_{j,m,\bar m}$
by an amount $w$ will be denoted as $\Phi^{w}_{j,m,\bar m}$
whose explicit form is
\eqn\aan{
\Phi^w_{j,m,\bar m}=e^{\sqrt{\frac{2}{n+2}}\big[(m+\frac{n+2}{2}w)X_3
+(\bar m+\frac{n+2}{2}w)\bar X_3\big]} V_{j,m,\bar m}
~.}
We will see in a moment that the spectral flow operation
appears naturally as part of the topological twist.

In any case, given the above primary fields
one can easily construct primary fields of
the supersymmetric coset by simply demanding
that the total $U(1)$-gauging charge $Q_g$ is zero.
For example, one can check that the primary
fields
\eqn\aao{
e^{i(sH-\bar s \bar H)}\Phi_{j,m,\bar m}
e^{i\sqrt{\frac{2}{n}}\big[(s+m)X-(\bar s+\bar m)\bar X\big]}
}
satisfy this condition and are therefore allowed primary fields
of the supersymmetric coset.

We are now ready to consider the topologically
twisted theory.
We define the A-model topological twist
of the $\NN=2$ coset by the usual twist of the worldsheet stress
tensor\foot{It is not difficult
to consider also the B-model topological twist,
but this will not be done explicitly in this paper.}
\eqn\aah{
T\rightarrow T+\frac{1}{2}\d J'_R~, ~ ~
\bar T\rightarrow \bar T-\frac{1}{2}\bar{\d} {\bar J}'_R
~.}
Notice that, as in \TakayanagiYB, we choose to define the topological twist
with the use of the gauge equivalent $U(1)_R$ current
$J'_R$. One can check that with these conventions the
BRST cohomology of the topologically twisted theory will consist
of $(c,a)$ primaries of the $\NN=2$ coset CFT.
By definition, the $(c,a)$ primary fields
are NS-sector primary fields $\OO_{NS}$
satisfying the constraints
\eqn\aap{
\oint dz ~ G^+(z) \cdot \OO_{NS} (0)=0~, ~ ~
\oint d\bar z ~ \bar G^-(\bar z)\cdot \OO_{NS}(0)=0
~.}

Some of these primaries (in fact the ones that will be
relevant for this paper) can be determined
simply by setting $s=\bar s=0$, $m=\bar m=j$ in
\aao. For later convenience, we denote these fields as
\eqn\aar{
V^{NS}_{-\frac{2j+1}{2n}}=\Phi_{j,j,j}
e^{i\sqrt{\frac{2}{n}}j(X-\bar X)}
~.}
They have scaling dimensions
$\Delta=\frac{q_R}{2}=-\frac{j}{n}$,
$\bar \Delta=-\frac{\bar q_R}{2}=-\frac{j}{n}$
and vanishing momentum in the angular direction
of the cigar.

An equivalent representation of the $(c,a)$ primary
fields can be given in the Ramond sector as R-sector ground
states \LercheUY. The corresponding vertex operators
will be denoted as $V^R_{-\frac{2j+1}{2n}}$. They can be obtained from
the NS-sector vertex operators \aar\ by $(-\frac{1}{2},\frac{1}{2})$
$\NN=2$ spectral flow (the relevant conventions are summarized
in appendix A)
\eqn\aas{
V^R_{-\frac{2j+1}{2n}}=e^{-\frac{i}{2}\sqrt{\frac{n+2}{n}}(-X_R+\bar X_R)}
V^{NS}_{-\frac{2j+1}{2n}}=
e^{\frac{i}{2}(H+\bar H)}\Phi_{j,j,j}
e^{i\sqrt{\frac{2}{n}}\big(j+\frac{1}{2}\big)(X-\bar X)}
~.}
The canonically normalized boson $X_R$ appearing
in the first equality is defined so that
\eqn\aasa{
J_R(z)=-i\sqrt{\frac{n+2}{n}}\d X_R(z)
~.}

Going in the opposite direction,
physical states of the topologically twisted coset
in the NS sector can be obtained from the R-sector ground states
by a $(\frac{1}{2},-\frac{1}{2})$ $\NN=2$ spectral flow
transformation, but now according to \aah\
we define the $\NN=2$ spectral flow transformation
in terms of the modified boson $X'_R$.
Applying this operation to the R-ground states \aas\
gives the vertex operators
\eqn\aat{
V_{-\frac{2j+1}{2n}}=e^{-\frac{i}{2}\sqrt{\frac{n+2}{n}}(X'_R-\bar X'_R)}
V^{R}_{-\frac{2j+1}{2n}}=
e^{-i(H+\bar H)}\Phi^{w=1}_{j,j,j}e^{i\sqrt{\frac{2}{n}}(j+\frac{n}{2})(X-\bar X)}
~.}
Note that with this prescription
part of the topological twist is
an $SL(2,\IR)$ spectral flow operation, since $X'_R$ contains
a term proportional to $X_3$.

We conclude this subsection with a few
additional comments on certain features of the topologically
twisted coset theory:
\item{(1)} In what follows we consider a
$\ZZ_n$ orbifold projection of the topologically twisted
coset $SL(2,\IR)_n/U(1)$. For this it is necessary to
have $n\in \IZ_+$. The $\IZ_n$ projection acts
by translation $\frac{2\pi}{\sqrt{n/2}}$ on the boson $X$
and reduces the asymptotic radius of the cigar from
$R=\sqrt{2n} \rightarrow \sqrt{\frac{2}{n}}$ (in $\alpha'=2$ units).
\item{(2)} The quantum numbers $j$ in \aat\ are real
and have been argued \TakayanagiYB\ to satisfy the constraint
$2j+1\in n\IZ$. This is consistent with the above orbifold and
can be deduced by demanding that the vertex operators $V_{-\frac{2j+1}{2n}}$
are mutually local with the $SU(2)$ $K^-$ operator.
Further details can be found in \TakayanagiYB.
\item{(3)} In standard discussions of the $SL(2,\IR)$ WZW model
or its $U(1)$ coset, one demands that normalizable representations
respect the unitarity bound $-(n+1)/2<j<-1/2$. This bound will be dropped in the
topologically twisted theory, since we focus mostly on non-normalizable
representations.
\item{(4)} The topologically twisted coset exhibits many more
physical states, which can be formulated most easily in
the Wakimoto free field representation of $SL(2,\IR)$
\refs{\MukhiZB,\TakayanagiYB}. For each of these states
there is a corresponding state in the dual bosonic string.
We described some of these states here corresponding to
ghost number $1$ tachyons, but there are
more at this ghost number and ghost numbers $0$ or $2$
and arrange themselves in $SU(2)$ multiplets.
An extensive list appears in \refs{\MukhiZB,\TakayanagiYB}.
In this paper we focus on the physical states \aat, since
they have a simple formulation without any reference to the
Wakimoto representation.
\item{(5)} It is well-known that the $\NN=2$ coset $SL(2)_n/U(1)$
is mirror dual to the $\NN=2$ Liouville theory \HoriAX. It would be
interesting to consider the topologically twisted $\NN=2$ Liouville theory and
repeat the analysis of \refs{\MukhiZB,\TakayanagiYB}
directly in this context to obtain
the non-trivial BRST cohomology of the $c\leq 1$ non-minimal
bosonic string. This seems to be
complicated by the absence of an obvious analogue of the
Wakimoto representation.

\subsec{The non-minimal $c\leq 1$ bosonic string}

The non-minimal $c\leq 1$ bosonic string was first
introduced in \TakayanagiYR. On the worldsheet level,
this theory consists of a time-like linear dilaton theory
with slope $q=\frac{1}{b}-b$ and the standard
space-like Liouville theory with linear dilaton slope
$Q=b+\frac{1}{b}$. In our conventions,
the worldsheet action takes the form
\eqn\aba{
\SS=\frac{1}{4\pi}\int d^2 z ~ \big(-\d X_0 \bar \d X_0+
\frac{q}{2\sqrt 2} X_0 \RR^{(2)}+
\d \phi \bar \d \phi+\frac{Q}{2\sqrt 2} \phi \RR^{(2)}+
2\pi \mu_L e^{\sqrt 2 b\phi}\big)
~}
and the total central charge is
\eqn\abc{
c_{tot}=c_{X_0}+c_{\phi}=1-6q^2+1+6Q^2=26
~.}
This theory is a natural extension of the $c=1$ non-critical
string, where $b=1$ and the time-like boson has vanishing
linear dilaton slope $q$. In general, the parameter $b$
is restricted to lie within the range
$0\leq b\leq 1$. The second condition comes
from the Seiberg bound on the Liouville potential $e^{\sqrt 2 b\phi}$.

The primary fields of the non-minimal bosonic string are
\eqn\abd{
\VV_{\alpha,\beta}=e^{\sqrt 2(\alpha X_0+\beta \phi)}
}
and have scaling dimensions
\eqn\abe{
\Delta(\VV_{\alpha,\beta})=-\alpha(q-\alpha)+\beta(Q-\beta)
~.}
Notice that the string coupling is
\eqn\abf{
g_s=e^{\frac{1}{\sqrt 2}(qX_0+Q\phi)}
~}
and the system will get strongly coupled at late times,
but an appropriate Lorentz transformation \TakayanagiYR\
can be used to recast this theory as an ordinary linear dilaton
vacuum perturbed by a time-dependent Liouville potential.
A matrix model formulation of this theory has been discussed
in \TakayanagiYR.

\subsec{The correspondence}

For integer numbers $n$
it has been argued \TakayanagiYB\ that the non-minimal
$c=1-6(n-1)^2/n$ string at radius $R_{X_0}=\sqrt{2n}$
is equivalent to the topologically twisted
A-model orbifold $\bigg(\frac{SL(2,\IR)_{n+2}}{U(1)}\bigg)/\IZ_{n}$.
The explicit mapping of states can be found in \TakayanagiYB.
In particular, the vertex operators that map to \aat\
are the bosonic string tachyons
\eqn\abg{
c\bar c ~\YY_s=c\bar c ~ \VV_{\alpha_s-\frac{1}{\sqrt n},\alpha_s}
~,}
where $c$ is the usual $c$ ghost of the bosonic string,
$\alpha_s$ is
\eqn\abi{
\alpha_s=\frac{1}{\sqrt n}\bigg(\frac{1}{2}-ns\bigg)+\frac{\sqrt n}{2}
}
and the relation between $s$ and the quantum number $j$ appearing
in \aat\ is $s=-\frac{2j+1}{2n}$.

Here it is perhaps useful to recall this mapping of states
in the more familiar $c=1$ context using the more standard notation
of \WittenYJ. The topological string vertex operators
$V_s$ ($s=-j-\frac{1}{2}\in \frac{\IZ}{2}$) in \aat\ correspond in the
$c=1$ bosonic string to the ghost number one
tachyons
\eqn\abj{\eqalign{
Y^+_{s,-s}&=c\bar c ~ e^{-i\sqrt 2 s X_0}e^{\sqrt 2 (1-s)\phi}~,
~ ~ s\geq 0~,
\cr
Y^-_{s,s}&=c\bar c ~ e^{i\sqrt 2 s X_0}e^{\sqrt 2 (1+s)\phi}~,
~ ~ s\geq 0
~.}}
The two indices $(s,\pm s)$ are $SU(2)_L\times SU(2)_R$ labels.
In addition, there is a conjugate class of topological string vertex operators
$\tilde V_s$, which will not be discussed explicitly here, that
correspond to the $c=1$ bosonic string tachyons
\eqn\abk{\eqalign{
Y^+_{s,s}&=c\bar c ~ e^{i\sqrt 2 s X_0}e^{\sqrt 2 (1-s)\phi}~,
~ ~ s\geq 0~,
\cr
Y^-_{s,-s}&=c\bar c ~ e^{-i\sqrt 2 s X_0}e^{\sqrt 2 (1+s)\phi}~,
~ ~ s\geq 0
~.}}
The vertex operators $Y^+_{s,\pm s}$ appear also
in \DiFrancescoUD\ in another notation
\eqn\abl{
T_k=c\bar c ~ e^{ikX_0}e^{(-|k|+\sqrt 2)\phi}
}
with $k=-\sqrt 2 s$, $s\in \frac{\IZ}{2}$.
The absolute value $|k|$ in the exponent
is chosen in order to satisfy the Seiberg bound.
$N$-point functions on the sphere of these vertex
operators have been computed in
\refs{\DijkgraafDJ\LosevTT-\EguchiXX,\MukhiZB,\TakayanagiYB}
using topological LG techniques and
in \refs{\DiFrancescoUD,\TakayanagiYR}
using the worldsheet formulation of the bosonic string.
In this paper we compute directly in
the $SL(2,\IR)/U(1)$ coset.

\newsec{The Stoyanovsky-Ribault-Teschner map}

Recently the authors of \RibaultWP\
formulated a very precise map between
sphere correlation functions in the $SL(2,\IC)/SU(2)$
WZW model and correlation functions in Liouville field
theory. The proof of this map is based on a relation
between the KZ and BPZ systems of partial differential
equations in the two theories and provides the extension
of a similar observation by Stoyanovsky in \StoyanovskyPG\
for $SU(2)$ WZW models.

The basic formula that was obtained in
\RibaultWP\ provides a map between
winding number conserving $N$-point
functions in the $SL(2,\IC)/SU(2)$ WZW model at level $k$
and $(2N-2)$-point functions in Liouville field theory
with linear dilaton slope $Q=b+b^{-1}$ and $b^2=\frac{1}{k-2}$.
More explicitly,
\eqn\baa{\eqalign{
\bigg\langle \prod^N_{i=1} \Phi_{j_i,m_i,\bar m_i}(z_i,\bar z_i)\bigg\rangle&=
\prod_{i=1}^N \NN^{j_i}_{m_i,\bar m_i}
\delta\bigg(\sum_{\ell=1}^N m_{\ell}\bigg) \delta\bigg(\sum_{\ell=1}^N \bar m_{\ell}\bigg)
\cr
&\prod_{a=1}^{N-2} \int d^2 y_a ~ \FF_k(\{z_i,\bar z_i\};\{y_a,\bar y_a\})
\bigg\langle \VV_{\alpha_i}(z_i,\bar z_i)\VV_{-\frac{1}{2b}}(y_a,\bar y_a)\bigg\rangle
~,}}
where
\eqn\bab{\eqalign{
\FF_k(\{z_i,\bar z_i\};\{y_a,\bar y_a\})&=\frac{2\pi^3 b}{\pi^{2N}(N-2)!}
\prod_{i<j\leq N} (z_i-z_j)^{m_i+m_j+\frac{k}{2}}
(\bar z_i-\bar z_j)^{\bar m_i +\bar m_j+\frac{k}{2}}
\cr
&\prod_{a<b\leq N-2} |y_a-y_b|^k
\prod_{r=1}^N \prod_{c=1}^{N-2} (z_r-y_c)^{-m_r-\frac{k}{2}}
(\bar z_r-\bar y_c)^{-\bar m_r-\frac{k}{2}}
}}
and
\eqn\bac{
\NN^j_{m,\bar m}=\frac{\Gamma(-j+m)}{\Gamma(1+j-\bar m)}
~.}
Following the notation of the previous section,
the Liouville field theory vertex operators
$\VV_{\alpha}$ reduce to the exponential fields
$e^{\sqrt 2 \alpha \phi}$ in the classical limit $b\rightarrow 0$.
The SRT map involves $N-2$ degenerate vertex operators
$\VV_{-\frac{1}{2b}}$ and $N$ vertex operators
with
\eqn\bad{
\alpha_i=bj_i+b+\frac{1}{2b}~, ~ ~ 1\leq i\leq N
~,}
which are in one-to-one
correspondence with the $SL(2,\IC)/SU(2)$ primary fields.
In these expressions the $SL(2,\IC)/SU(2)$ quantum numbers $(j,m,\bar m)$
are restricted to the set of values
\eqn\bae{
j=-\frac{1}{2}+is~, ~ ~ m=\frac{n+ip}{2}, ~ \bar m=\frac{-n+ip}{2},
~ ~ n\in \IZ, ~ s,p\in \IR
~.}
The corresponding primary fields belong to the continuous series
and cover the full physical spectrum of the $SL(2,\IC)/SU(2)$
model. In order to apply this map to $SL(2,\IR)$ correlation
functions we need to perform an analytic continuation, both
in the $SL(2,\IC)/SU(2)$ WZW model and in Liouville field theory.
In general, this analytic continuation is believed to hold
and gives sensible results as we see below.

In addition, as we summarized above,
the physical spectrum of string theory on $SL(2,\IR)$ also
includes spectral flowed representations. Correlation functions
involving such primary fields are not expected to obey the
KZ equations, but there is a natural generalization
of the KZ equations \RibaultMS\ that allows for the inclusion of
spectral flow. Then, one can show that with the appropriate
modifications the SRT map with spectral flowed primary fields
takes the form \RibaultMS\foot{Compared to the conventions
of \RibaultMS\ we have $m_{ours}=-m_{there}$,
$\bar m_{ours}=-\bar m_{there}$
and $w_{ours}=-w_{there}$.}
\eqn\baf{\eqalign{
\bigg\langle \prod_{i=1}^N \Phi^{w_i}_{j_i,m_i,\bar m_i}(z_i,\bar z_i)\bigg\rangle
_{\sum w=r\geq 0}&=
\prod_{i=1}^N \NN^{j_i}_{m_i,\bar m_i}
\delta^{(2)}\bigg(\sum_{\ell=1}^N m_{\ell}+\frac{k}{2} r\bigg)
\cr
&\prod_{a=1}^{N-2-r} \int d^2 y_a ~ \tilde{\FF}_k(\{z_i,\bar z_i\};\{y_a,\bar y_a\})
\bigg\langle \VV_{\alpha_i}(z_i,\bar z_i)\VV_{-\frac{1}{2b}}(y_a,\bar y_a)\bigg\rangle
~,}}
where $c_k$ is a $k$-dependent constant and
\eqn\bag{\eqalign{
\tilde{\FF}_k(\{z_i,\bar z_i\};\{y_a,\bar y_a\})&=\frac{2\pi^{3-2N}b c_k^r}{(N-2-r)!}
\prod_{i<j\leq N} (z_i-z_j)^{\beta_{ij}}
(\bar z_i-\bar z_j)^{\bar{\beta}_{ij}}
\cr
&\prod_{a<b\leq N-2-r}|y_a-y_b|^k
\prod_{r=1}^N \prod_{c=1}^{N-2-r} (z_r-y_c)^{-m_r-\frac{k}{2}}
(\bar z_r-\bar y_c)^{-\bar m_r-\frac{k}{2}}
~}}
with
\eqn\bah{\eqalign{
\beta_{ij}&=\frac{k}{2}+m_i+m_j-\frac{k}{2}w_i w_j-w_i m_j-w_j m_i ~,
\cr
\bar \beta_{ij}&=\frac{k}{2}+\bar m_i+\bar m_j-\frac{k}{2}\bar w_i \bar w_j
-\bar w_i \bar m_j-\bar w_j \bar m_i
~.}}

As we have already seen in section 2, part of the topological
twist involves spectral flow in $SL(2,\IR)$. Hence,
the map \baf\ will be directly relevant for the topological
string computation of the next section.

\newsec{$N$-point functions}

This section contains the main observation of this paper.
Using the SRT map, we compute $N$-point functions
of the topological string vertex operators \aat\
and relate them to $N$-point functions in the
corresponding non-minimal $c\leq 1$ bosonic string.
We begin with a brief review of a well-known
prescription that allows us to recast the topological
string correlation functions as correlation functions
in the untwisted CFT.

\subsec{General comments}

With each physical state $|\OO_{\alpha}\rangle$ in a
topologically twisted (A-model) theory one can associate a complete
superfield
\eqn\caa{
\OO_{\alpha}=\OO^{(0)}_{\alpha}+\theta \OO^{(1,0)}_{\alpha}
+\bar{\theta} \OO^{(0,1)}_{\alpha}+
\theta \bar{\theta}\OO^{(1,1)}_{\alpha}
~,}
or one can write in the NS-sector
\eqn\cab{
\OO^{(1,0)}_{\alpha}=G^-_{-1/2}\cdot \OO^{(0)}_{\alpha}, ~ ~
\OO^{(0,1)}_{\alpha}=\bar G^+_{-1/2}\cdot \OO^{(0)}_{\alpha}, ~ ~
\OO^{(1,1)}_{\alpha}=G^-_{-1/2}\bar G^+_{-1/2} \cdot \OO^{(0)}_{\alpha}
~.}
Accordingly, there are four topological string observables associated to these
components:
\eqn\cac{
\OO^{(0)}_{\alpha}~, ~ ~ \oint dz~ \OO^{(1,0)}_{\alpha}~, ~ ~
\oint d\bar z ~ \OO^{(0,1)}_{\alpha}~, ~ ~
\int d^2 z~ \OO^{(1,1)}_{\alpha}
~.}
For example, for the topological vertex operators
$V_{-\frac{2j+1}{2n}}$ appearing in \aat\ we get
\eqn\caca{
V^{(1,0)}_{-\frac{2j+1}{2n}}\equiv G^-_{-1/2}\cdot V_{-\frac{2j+1}{2n}} =
2\sqrt{\frac{2}{n}}j ~e^{-i\bar H}\Phi^{w=1}_{j,j-1,j}
e^{i\sqrt{\frac{2}{n}}(j+\frac{n}{2})(X-\bar X)}
~.}
Analogous results apply also to the $(0,1)$ and $(1,1)$ forms
of this vertex operator.

In general, correlation functions
$\langle \prod_{i} \OO_{\alpha_i}^{(0)}\rangle$
of the $0$-form operators $\OO^{(0)}_{\alpha_i}$ can be
determined easily by using factorization in terms of the
three-point function \DijkgraafDJ
\eqn\cad{
c_{\alpha_1 \alpha_2 \alpha_2}=
\langle \OO^{(0)}_{\alpha_1} \OO^{(0)}_{\alpha_2}
\OO^{(0)}_{\alpha_3}\rangle
~.}
In a similar fashion, one can show that the more general
correlation function of the type\foot{We will not consider
correlation functions involving the observables
$\oint dz~ \OO^{(1,0)}_{\alpha}$ or
$\oint d\bar z ~ \OO^{(0,1)}_{\alpha}$ in this paper.}
\eqn\cae{
\bigg\langle \prod_{i=1}^N \OO^{(0)}_{\alpha_i}
\prod_{\ell=1}^M \int d^2 z_{\ell} ~ \OO^{(1,1)}_{\alpha_{\ell}}\bigg\rangle
~}
can be recast in terms of the perturbed three-point
function
\eqn\caf{
c_{\alpha_1 \alpha_2 \alpha_3}(t)=
\bigg\langle \OO^{(0)}_{\alpha_1} \OO^{(0)}_{\alpha_2}
\OO^{(0)}_{\alpha_3} e^{\sum_n t_n \int d^2 z~ \OO^{(1,1)}_{\alpha_n}}
\bigg\rangle
~}
or in terms of correlation functions of the form
\eqn\cag{
\FF_{\alpha_1,...,\alpha_N}\equiv
\bigg\langle \OO^{(0)}_{\alpha_1} \OO^{(0)}_{\alpha_2}
\OO^{(0)}_{\alpha_3} \prod_{\ell=4}^N \int d^2 z_{\ell} ~
\OO^{(1,1)}_{\alpha_{\ell}}\bigg\rangle
~.}
It is precisely this last type of correlation functions
that we want to determine for the vertex operators
\aat. This can be achieved in the
untwisted $\NN=2$ coset CFT in the following way.

The topologically twisted theory exhibits a $U(1)_R$
anomaly (or a $U(1)_R$ background charge) of -$\frac{c}{3}$
on the sphere. In the original untwisted $\NN=2$ superconformal
theory, the $U(1)_R$ current is not anomalous and the corresponding
charge is conserved. Therefore, in order to compute the
correlation function \cag\ in the untwisted theory
we need to insert the background charge by hand. This
can be achieved in many different ways, but the one
we choose to employ here is the following. We can split
the background charge into two pieces and introduce
the $(\frac{1}{2},-\frac{1}{2})$ spectral-flow operator\foot{The opposite
signs for the left- and right-movers here have to do
with the particulars of the A-model topological
twist described in section 2.}
\eqn\cah{
\mu(z,\bar z)=e^{-\frac{i}{2}\sqrt{\frac{c}{3}}
(X'_R(z)-X'_R(\bar z))}
}
at two distinct points, say $\xi_1$ and $\xi_2$.
Then, we can take $\xi_1$ and $\xi_2$ respectively to coincide with
the insertions of the first two vertex operators
$\OO^{(0)}_{\alpha_1}$ and $\OO^{(0)}_{\alpha_2}$
and this effectively converts them into vertex operators
in the R-sector (see, for example, \aas).\foot{From now on,
we consider the $0$-form vertex operators
exclusively in the NS-sector (they are $e.g.$ of
the form \aat).}
For simplicity we may set $\xi_1=0$, $\xi_2=\xi$
and later send $\xi \rightarrow \infty$.
At the end, we can re-express the $N$-point function
\cag\ in terms of the following correlation function
in the untwisted $\NN=2$ theory
\eqn\cai{\eqalign{
&\FF_{\alpha_1,...,\alpha_N}=
\cr
&\lim_{\xi \rightarrow \infty}|\xi|^{A_{123}}
\bigg\langle \OO^{(0)R}_{\alpha_1}(0)\OO^{(0)R}_{\alpha_2}(\xi,\bar \xi)
\OO^{(0)}_{\alpha_3}(1)\prod_{\ell=4}^N \int d^2 z_{\ell}
|z_{\ell}|^{q_{\alpha_{\ell}}-1}|z_{\ell}-\xi|^{q_{\alpha_{\ell}}-1}
\OO^{(1,1)}_{\alpha_{\ell}}(z_{\ell},\bar z_{\ell})
\bigg\rangle
~,}}
where $q_{\alpha_{\ell}}$ denotes the $U(1)_R$ charge of
the vertex operators $\OO^{(0)}_{\alpha_{\ell}}$
and the constant $A_{123}$ is
\eqn\caj{
A_{123}=q_{\alpha_1}+q_{\alpha_2}+q_{\alpha_3}-\frac{c}{6}
~.}
The extra insertions
$|z|^{q_{\alpha_{\ell}}-1}|z-\xi|^{q_{\alpha_{\ell}}-1}$
and the factor $|\xi|^{A_{123}}$ are
needed in order to isolate the dimensionless part of
the $N$-point function in the untwisted theory, which is
precisely the one that coincides with the topological
string amplitude. For more details on this prescription
we refer the readers to \refs{\WittenZZ,\WarnerZH}.

\subsec{Computing $N$-point functions}

We now focus on the topologically
twisted coset CFT $SL(2,\IR)_n/U(1)$ at level $n$.
We want to compute the scattering amplitude
of $N$ vertex operators of the form \aat. According to
the general prescription \cai\ in the $untwisted$ coset theory
we should compute the amplitude
\eqn\daa{
\FF^n_V(\{s_i\};N)=
\lim_{\xi\rightarrow \infty} |\xi|^{A_{123}}\prod_{\ell=4}^N
\int d^2 z_{\ell} |z_{\ell}|^{q_{s_{\ell}}-1}|\xi-z_{\ell}|^{q_{s_{\ell}}-1}
\bigg\langle V^R_{s_1}(0)V^R_{s_2}(\xi,\bar \xi) V_{s_3}(1)
V^{(1,1)}_{s_{\ell}}(z_{\ell},\bar z_{\ell})\bigg\rangle
~.}
In this case, the constant $A_{123}$ is
\eqn\dab{
A_{123}=-\frac{3}{2}+2s_1+2s_2+2s_3+\frac{n+2}{n}
}
and the $U(1)_R$ charges
\eqn\dac{
q_{s_{\ell}}=1+2s_{\ell}-\frac{n-1}{n}
~.}
For quick reference, we also summarize the explicit
form of the vertex operators appearing in \daa
\eqn\dad{
V^R_s=e^{\frac{i}{2}(H+\bar H)}
\Phi_{-\frac{1}{2}-ns,-\frac{1}{2}-ns,-\frac{1}{2}-ns}
~e^{-i\sqrt{2n}s(X-\bar X)}
~,}
\eqn\dae{
V_s=e^{-i(H+\bar H)}
\Phi^{w=1}_{-\frac{1}{2}-ns,-\frac{1}{2}-ns,-\frac{1}{2}-ns}
~e^{i\sqrt{\frac{2}{n}}(-ns+\frac{n-1}{2})(X-\bar X)}
~,}
\eqn\daf{
V_s^{(1,1)}=\Phi^{w=1}_{-\frac{1}{2}-ns,-\frac{3}{2}-ns,-\frac{3}{2}-ns}
~e^{i\sqrt{\frac{2}{n}}(-ns+\frac{n-1}{2})(X-\bar X)}
~.}
In all cases the quantum number $s$ is a half-integer.
Also, notice that in \daf\ we have defined the normalization
of the $(1,1)$-form vertex operators to be such that\foot{In
the right-moving sector $\bar G^+=\sqrt{\frac{2}{n}}\bar \psi \bar J^-$ and
$\bar G^-=\sqrt{\frac{2}{n}}\bar \psi^{\dagger} \bar J^+$.}
\eqn\dafa{
V_s^{(1,1)}=\frac{n}{2(1+2ns)^2}~ G^-_{-1/2} \bar G^+_{-1/2} \cdot V_s
~.}
This is needed in order to obtain a correlation function
$\FF^n_V(\{s_i\};N)$ that is fully symmetric under the
interchange of the labels $s_i$.

For starters, let us consider the selection rules in \daa\ coming
from the different $U(1)$'s. The $H$-momentum conservation
condition is automatically satisfied and the conservation of
the $X$-momentum gives
\eqn\dag{
n\sum_{i=1}^N s_i=(N-2)\frac{n-1}{2}
~.}
This implies that the $J^3$-momenta satisfy the equations
\eqn\dai{
\sum_{i=1}^N m_i+(N-2)\frac{n+2}{2}=
\sum_{i=1}^N \bar m_i+(N-2)\frac{n+2}{2}=0
~.}

It is now straightforward to compute the correlation
function \daa\ with the use of the SRT map \baf.
The $U(1)$ amplitudes are trivial and
the $SL(2,\IR)$ correlation functions can be recast as
correlation functions in Liouville field theory.
There are certain things to notice in this computation.
First, in order to use the SRT map we must perform an analytic
continuation to the quantum numbers of the discrete representations
appearing in the topological string. Secondly, notice that
the amplitude \daa\ involves $N-2$ $SL(2,\IR)$ vertex operators,
each one having $w=1$. Hence, the amplitude \daa\
reduces to an $SL(2,\IR)$ $N$-point function with maximal
spectral-flow number violation. With the use of the SRT map \baf\
this amplitude will be recast in terms of an $N$-point function
in Liouville field theory. Finally, notice that
the $J^3$ selection rules appearing in
\baf\ are precisely the ones derived from the conservation
of the $X$-momentum \dai. Then, with a small amount of algebra
we find that
\eqn\daj{\eqalign{
\FF^n_{V}(\{s_i\};N)&=(-)^{N+1}\Gamma(0)^N~
\frac{2\pi^{3-2N} c^{N-2}_{n+2}}{\sqrt n}~
\delta\bigg(n\sum_{i=1}^N s_i-(N-2)\frac{n-1}{2}\bigg)
\cr
&\lim_{\xi\rightarrow \infty}
|\xi|^{4+\frac{4}{n}(-ns_2+\frac{n-1}{2})(-ns_1-ns_3+n-1)}
\prod_{\ell=4}^N\int d^2 z_{\ell} ~
|z_{\ell}|^{\frac{4}{n}(-ns_1+\frac{n-1}{2})(-ns_{\ell}+\frac{n-1}{2})}
\cr
&|1-z_{\ell}|^{\frac{4}{n}(-ns_3+\frac{n-1}{2})(-ns_{\ell}+\frac{n-1}{2})}
|\xi-z_{\ell}|^{\frac{4}{n}(-ns_2+\frac{n-1}{2})(-ns_{\ell}+\frac{n-1}{2})}
\cr
&\bigg\langle \VV_{\alpha_1}(0)\VV_{\alpha_2}(\xi,\bar \xi)
\VV_{\alpha_3}(1)\VV_{\alpha_{\ell}}(z_{\ell},\bar z_{\ell})
\bigg\rangle
~,}}
where the indices $\alpha_{s_i}$ (for $i=1,2,...,N$) are given by
eq.\ \abi\ and the Liouville field theory
has linear dilaton slope $Q=\sqrt n+\frac{1}{\sqrt n}$.
Already at this stage, simply from the conservation laws
and the value of the linear dilaton slope $Q$ we can see
quite clearly the emergence of the expected non-minimal bosonic string.

Indeed, if the topologically twisted coset $SL(2,\IR)_n/U(1)$
is equivalent to the non-minimal bosonic string
of section 2 with $b=\frac{1}{\sqrt n}$, then we should
be able to reproduce the above computation in
bosonic string theory. In other words, we should be able to show that
there is agreement between \daj\ and the $N$-point
function
\eqn\dba{
\FF_{\YY}^{c\leq 1}(\{s_i\};N)=
\lim_{\xi\rightarrow \infty} \prod_{\ell=4}^N \int d^2 z_{\ell}~
\bigg\langle c\bar c ~ \YY_{s_1}(0) c\bar c ~ \YY_{s_2}(\xi,\bar \xi)
c\bar c ~ \YY_{s_3}(1) \YY_{s_{\ell}}(z_{\ell},\bar z_{\ell})
\bigg\rangle
~.}
Straightforward computation of the $U(1)$ parts of this amplitude
yields the following relation
\eqn\dbb{
\mathboxit{\FF^n_{V}(\{s_i\};N)=(-)^{N+1}\Gamma(0)^N ~
\frac{2\pi^{3-2N} c^{N-2}_{n+2}}{\sqrt n}~
\FF_{\YY}^{c\leq 1}(\{s_i\};N)~.}
}
This formula is the main result of this paper.
We find that the topological string and bosonic
string computations agree up to an $n$ dependent
diverging factor. Rescaling the topological
string vertex operators with the extra
factor $-\frac{c_{n+2}}{\pi^2} \Gamma(0)$, $i.e.$ sending
\eqn\dbc{
V_s \rightarrow -\Gamma(0)\frac{c_{n+2}}{\pi^2} V_s
}
we conclude that the topological string and bosonic
string amplitudes agree up to the
factor
\eqn\dbd{
-\frac{2\pi^3}{\sqrt n ~c^2_{n+2}}
~.}
We would need to know explicitly
the possibly diverging constant $c_{n+2}$
in order to determine this factor. Agreement
up to a diverging constant factor was also noted
for the three-point functions in \TakayanagiYB.

\newsec{Conclusions}

In this paper we saw that on the level of
correlation functions on the sphere the SRT map
offers a very natural and straightforward way to go
from the topologically twisted $SL(2,\IR)/U(1)$ coset
to a corresponding non-critical bosonic string.
It is an interesting question whether we can make a
similar application of the SRT map to check
the bosonic/topological correspondence on the level
of more correlation functions involving, for example,
tachyons with general $SU(2)_L \times SU(2)_R$ labels.
But with this point aside, it is already interesting
that the set of correlation functions in \dbb\
provides us with a sufficiently large set of data
to reconstruct the correspondence
of \TakayanagiYB. To put it differently, if we did not
know about the non-minimal bosonic string of \TakayanagiYR\
and its relation to the topologically twisted coset,
we could have discovered it simply by using the SRT
map. This suggests that by using the SRT map
perhaps it will be possible to extend the
bosonic/topological equivalence to many more situations
where the $SL(2,\IR)/U(1)$ model is involved. A partial
list of interesting problems in this direction
is the following.

One possibility is to extend the equivalence
of \TakayanagiYB\ for levels $n$, which are not
integers. Rational values of $n$ are perhaps the
most natural and physically interesting extensions.
In this case, most of the subtleties have to do
with the $U(1)$ part of the non-minimal string and
the SRT map is not expected to play a crucial role
in this extension.

Another wide class of interesting examples involving
the supersymmetric coset $SL(2,\IR)/U(1)$
are those that are related to the $\NN=2$ topological
string in the vicinity of general
CY threefold singularities. According to \GiveonZM\
these theories are dual
(in a double-scaling limit) to
a topological string with target space of the general form
$SL(2,\IR)/U(1) \times ({\rm some}~ \NN=2~ {\rm CFT})$
and are relevant for the topological sector of
corresponding four-dimensional Little String Theories
(LST's) \refs{\AharonyKS,\KutasovUF}. It would be interesting to uncover
equivalent non-critical bosonic string theories
in these situations and see how they fit into the
general picture that anticipates these bosonic
theories as deformations of the $c=1$ string \VafaWI.
It would also be interesting to see if it is possible
to connect this story with the matrix model
statements of \AganagicQJ.

An obvious question is whether such bosonic/topological
string equivalences are particular to the $\NN=2$ topological
string, or whether more general string/string equivalences appear
also for other topological strings like the $\NN=2$
string \refs{\OoguriFP,\OoguriIE}.
It would certainly be worthwhile considering the
$\NN=2$ string on spaces of the form
$SL(2,\IR)/U(1) \times ({\rm some}~\NN=2~ {\rm CFT})$ \refs{\AharonyVK,\KonechnyDF},
which by the arguments of \GiveonZM\ are dual to $\NN=2$
strings in K3 singularities, and compute scattering
amplitudes using the SRT map.\foot{For a recent discussion
of scattering amplitudes in non-critical $\NN=2$ strings
see \TakayanagiYJ.} This computation would be relevant
to the physics of the topological sector of six-dimensional
LST's. The above issues are currently under investigation \nania.



\appendix{A}{$\NN=2$ and $SL(2,\IR)$ spectral flow conventions}

For quick reference we summarize in this appendix the definition of
the spectral flow transformation for the $\NN=2$ superconformal
algebra and the bosonic $SL(2,\IR)$ WZW model at level
$k$.

For the $\NN=2$ superconformal algebra with central charge $c$
the spectral flow transformation
by an amount $\theta$ (or $(\theta,\bar \theta)$ if we choose to include
both the left- and right-movers) is the $\NN=2$ superconformal
algebra transformation
\eqn\waa{\eqalign{
\tilde L_n&=L_n+\theta J_n+\theta^2 \frac{c}{6}\delta_{n,0}
\cr
\tilde G^{\pm}_n&=G^{\pm}_{n\pm \theta}
\cr
\tilde J_n&=J_n+\frac{c}{3} \theta
~.}}
On the level of primary fields this transformation can be achieved
by multiplying with the vertex operator $e^{-i\theta \sqrt{\frac{c}{3}}X_R}$,
where $X_R$ is a canonically normalized boson that bosonizes
the $U(1)_R$ current.

In a similar fashion, in $SL(2,\IR)$ at level $k$
the spectral flow operation with winding number
$w\in \IZ$ is defined by the current algebra automorphism\foot{Compared
to the conventions of Maldacena and Ooguri in \MaldacenaHW\
we have $w_{ours}=-w_{MO}$.}
\eqn\wab{
\tilde J_n^3=J_n^3-\frac{k}{2}w \delta_{n,0}~,
~ ~ \tilde J^{\pm}_n=J^{\pm}_{n\mp w}
~.}
On the level of Virasoro generators the spectral flow transformation
acts by the shift
\eqn\wab{
\tilde L_n=L_n-w J^3_n-\frac{k}{4}w^2\delta_{n,0}
~.}
On vertex operators it corresponds to multiplication
with the operator $e^{w \sqrt{\frac{k}{2}}X_3}$.
$X_3$ is the canonically normalized boson
that bosonizes the current $J^3$.

\listrefs
\end